\documentclass[fleqn,usenatbib]{mnras}

\usepackage{newtxtext,newtxmath}

\usepackage[T1]{fontenc}
\usepackage{graphicx}
\usepackage{hyperref}

\DeclareRobustCommand{\VAN}[3]{#2}
\let\VANthebibliography\thebibliography
\def\thebibliography{\DeclareRobustCommand{\VAN}[3]{##3}\VANthebibliography}

\usepackage{graphicx}	
\usepackage{amsmath}	
\usepackage{subcaption}

\newcommand{\kms}{$\,{\rm km/s}$}	
\newcommand{\kpc}{$\,{\rm kpc}$}	
\newcommand{\kpckms}{$\,{\rm kpc\,km/s}$}	
\newcommand{\KE}{$\,{\rm (kpc\,km/s)^2}$}	
\newcommand{\msun}{$\,{\rm M}_{\odot}$}
\newcommand{\gyr}{$\,{\rm Gyr}$}

\makeatletter
\DeclareRobustCommand{\HI}{%
  \mbox{H\check@mathfonts\fontsize\sf@size\z@\selectfont I}%
}
\makeatother

\makeatletter
\DeclareRobustCommand{\HII}{%
  \mbox{H\check@mathfonts\fontsize\sf@size\z@\selectfont II}%
}
\makeatother

\title[Retrograde stellar clusters in M33]{Unveiling M33’s Hidden Merger History: \\ A Potential Population of Star Clusters on Retrograde Orbits}

\author[Anguiano, Lewis \& Majewski]{
Borja Anguiano,$^{1,2}$\thanks{E-mail: banguiano@cefca.es}\thanks{Ramon y Cajal Fellow}
Geraint F. Lewis, $^{3}$ and
Steven R. Majewski$^{2}$
\\
$^{1}$Centro de Estudios de F\'isica del Cosmos de Arag\'on (CEFCA), Plaza San Juan 1, 44001, Teruel, Spain\\
$^{2}$Department of Astronomy, University of Virginia, Charlottesville, VA 22904, USA\\
$^{3}$Sydney Institute for Astronomy, School of Physics, A28, The University of Sydney, NSW 2006, Australia
}

\date{Accepted XXX. Received YYY; in original form ZZZ}

\pubyear{\the\year{2025}}

\begin{document}
\label{firstpage}
\pagerange{\pageref{firstpage}--\pageref{lastpage}}
\maketitle

\begin{abstract}
We report the discovery of a possible sub-population of stellar clusters that appear to follow retrograde orbits around the third largest galaxy in the Local Group, M33 (Triangulum). This spiral disk galaxy has apparently had a mostly quiescent existence, although recent discoveries, particularly of a pronounced warp in the gas and stellar disk, suggest that M33's relatively quiet past was interrupted at least once by a dynamical interaction with another galaxy. We suggest that this sub-population provides evidence of accretion of one or more dwarf galaxies in M33's history. We estimate a lower limit for the accreted halo virial mass of $M_{\rm vir} \sim (7 \pm 3) \times 10^{10}$\msun, accounting for about 10\% of the virial mass in the halo of M33 that has an accretion origin. We propose one of these accretion events as the source of the observed warp in M33's disk.
\end{abstract}




\section{Introduction}
The $\Lambda$ cold dark matter ($\Lambda$CDM) paradigm predicts that galaxies form hierarchically, growing  through the accretion of smaller systems \citep{springel+2006}. Clues to the accretion history of large galaxies has be inferred from the spatial, kinematic and chemical distribution of stellar clusters that would have also accreted with disrupting dwarf galaxies \citep{searle+1978,Brodie2006,Marin_Franch2009,kruijssen+2019}. The observational properties of these clusters typically span a wide range of ages, and their chemistry and kinematics retain signatures of some of the myriad processes that shaped their host galaxy \citep{Caldwell2011,Belokurov2024}. In the case of the Milky Way (MW) galaxy, there are cluster populations associated with major accretion events, e.g., the Sagittarius dwarf galaxy \citep{Law2010,Bellazzini2020} and Gaia-Sausage/Enceladus \citep{massari+2019,Limberg2022}, while for the Andromeda galaxy, a global alignment between clusters and streams has been reported \citep{McConnachie2018,Mackey2010,Wang2023}, suggesting a joint origin through accretion. These clusters offer the opportunity to trace the velocities of the tidal streams of the progenitor systems \citep{Huxor2011,Veljanoski2014}.

The Triangulum Galaxy (M33) is the third most massive object in the Local Group (LG), after Andromeda (M31) and the Milky Way \citep{vandenbergh2000}. Using deep optical and spectroscopic surveys, a stellar halo component of M33 has been determined \citep{chandar+2002,mcconnachie2006,Cullinane2023}, but there is no clear evidence for it possessing coherent stream-like substructures like those found in the other large LG galaxies \citep{Ferguson2007,Gilbert2022}. The existence of stellar clusters in the halo of M33 is well established \citep{sarajedini+2000, chandar+2006, beasley+2015, huxor+2009}, but the number is apparently small. \cite{cockcroft+2011} have suggested that some M33 halo clusters could potentially have been stripped away by M31 in a previous tidal interaction. The galaxies M33 and M31 are almost certainly gravitationally bound to one another \citep{Loeb2005}, hence M33 could show signatures of tidal perturbation. For example, M33 has evidence for a strong warp in neutral hydrogen \citep{Rogstad1976,Corbelli1997,Putman2009}, although some have concluded that the tidal force of M31 is unlikely to be sufficient to generate this warp \citep{Rogstad1976,Corbelli1997}. On the other hand, disturbed stellar distributions found outside M33’s \HI\ disk have been reported and suggested to have an origin either via tidal interaction with M31, if not from M33 accretion events \citep{Grossi2011}. Most definitively, the finding that M33 is enveloped by a prominent, diffuse stellar structure linked to M31 \citep{McConnachie2009} is persuasive evidence that the two systems are linked via a recent encounter. Yet, surprisingly, proper motions from the \emph{Gaia} mission show that M33 may be on its first infall into M31 \citep{vanderMarel2019}, which complicates the origin scenario for the extended feature. While our understanding of the structure of the M33 halo is not settled \citep[e.g.,][]{mcmonigal+2016,Gilbert2022,Ogami2024}, the position of a new dwarf satellite galaxy in the M33 halo seems to support the first infall scenario \citep{collins2024}.

\begin{figure*}
\centering
\begin{subfigure}[T]{0.48\linewidth}
  \includegraphics[width=\linewidth]{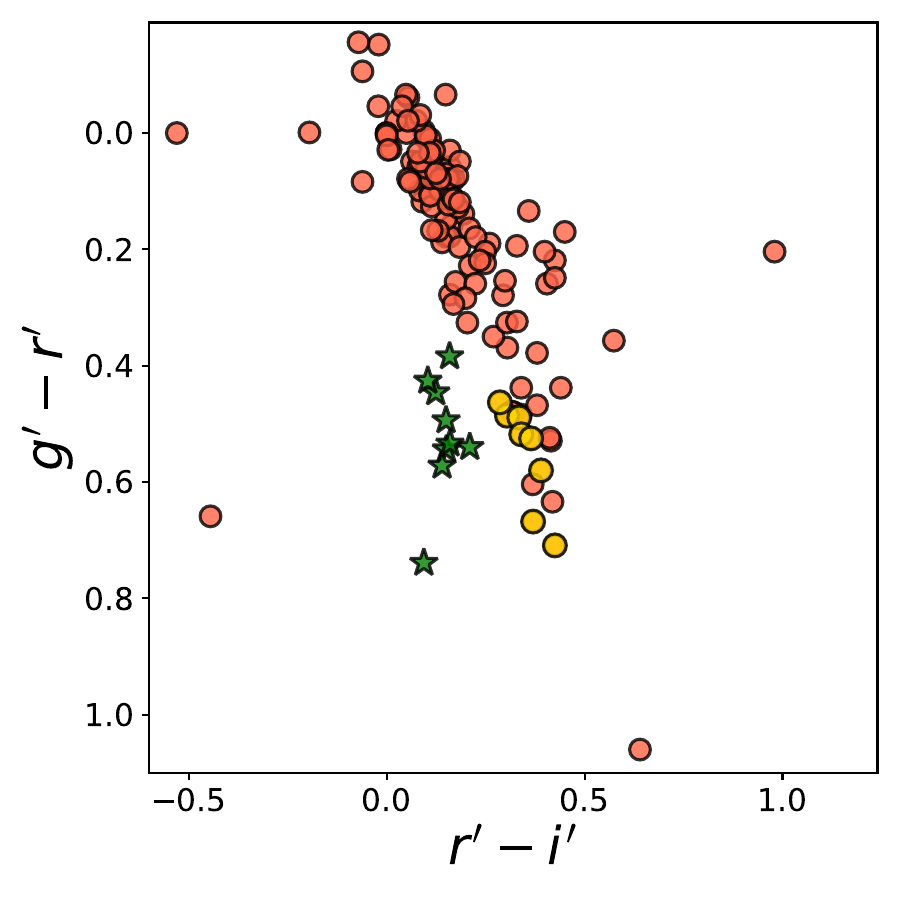}
\end{subfigure}%
\hspace{0.02\linewidth} 
\begin{subfigure}[T]{0.47\linewidth}
  \includegraphics[width=\linewidth]{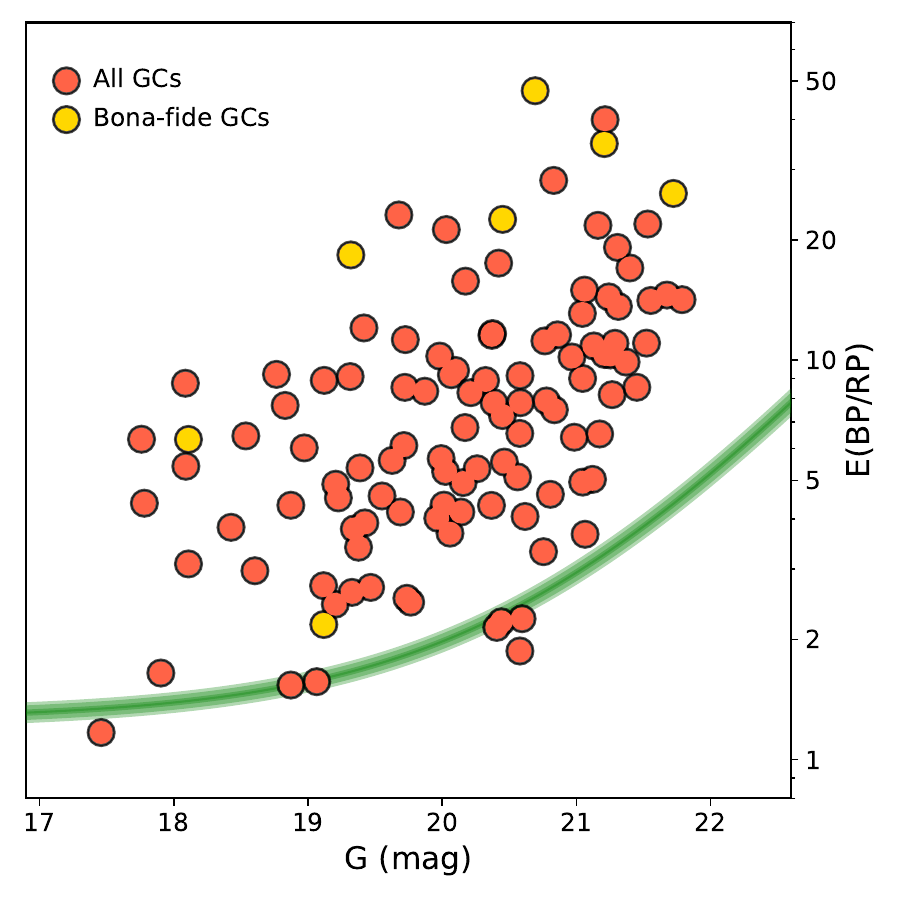}
\end{subfigure}
\caption{\emph{Left panel:} The ($r^{'}$ - $i^{'}$) - ($g^{'}$ - $r^{'}$) colour-colour diagram showing potential MW objects previously classified as M33 clusters (green stars) that we have removed according to the criteria described in Section \ref{sec:data}. Yellow circles indicate \emph{bona fide} globular clusters based on Hubble Space Telescope (HST) imaging. Orange circles represent the analysed cluster catalogue. {\emph Right panel:} {\emph Gaia} DR3 BP/RP excess factor as a function of \emph{Gaia} magnitude. The line represents the 3$\sigma$-above-the-mean foreground star BP/RP-excess factor value \citep{Hughes2021}. Objects falling below this line are removed from our sample as potential MW foreground stars. The figure shows seven {\emph bona fide} clusters that have BP/RP excess factor values, all of them clearly above the 3$\sigma$ limit line.}
\label{fig:figure1}
\end{figure*}

Globular cluster systems are powerful probes of galaxy formation and evolution \citep{Brodie2006}, especially through their kinematics, ages, and metallicity distributions, which sensitively trace the mass growth and minor merger history of the host galaxy \citep{forbes+2010,Law2010,Kruijssen2019}. Clusters on retrograde orbits, in particular, are key signatures of accretion \citep{Rodgers1984,Zinn1993}. The average age of halo globular clusters in M33 appears to be at least several gigayears younger than those in the Milky Way, M31, and other nearby galaxies \citep{Valenzuela2023}. Furthermore, the halo clusters in M33 formed over a much longer period than those in the Milky Way and M31 \citep{sarajedini+1998,chandar+2006,beasley+2015}. 

In this study, we present the discovery of a potentially accreted population of stellar clusters on retrograde orbits which we interpret as the signature of M33 accretion. The kinematic data for these stellar clusters is taken from spectroscopic samples available in the literature \citep{chandar+2002,sharina+2010,beasley+2015}. The layout of this paper is as follows: In Section~\ref{sec:data} we describe the data used in this study, while in Section~\ref{sec:targets} we discuss the kinematic properties of this sample together with the age-metallicity relation. We present our conclusions in Section~\ref{sec:conclusions}.

\begin{table*}
  \centering
  \caption{Retrograde clusters in M33 where $-Lz + 2\sigma_{Lz} < 0$ (see Section~\ref{sec:bulk}). In Notes, an entry of ``1'' indicates that the cluster has Str\"omgren photometry and \emph{Gaia} BP/RP excess factor values, while ``2'' indicates there is no Str\"omgren photometry for this cluster, however these clusters have an \emph{Gaia} BP/RP excess factor and they are marked as extended sources. For U77, the coordinate along the major axis, $X = -7.9 \times 10^{-5}$, is very small. This small value of $X$ significantly magnifies the velocity calculation and the uncertainty associated with it.}
  \begin{tabular}{cccccccc}
    \hline
    \textbf{RAJ2000} & \textbf{DECJ2000} & \textbf{$R$} & \textbf{$-L_z$} & \textbf{KE / 10$^{3}$}  & \textbf{[M/H]} & \textbf{Age} & \textbf{Notes, ID} \\
    degrees & degrees & \kpc & \kpckms & \KE & & \gyr  \\
    \hline
    \hline
    23.2768 & 30.6265 & 3.6 & --844 $\pm$ 240 & 27 $\pm$ 15 & -- & -- & 1 \\
    23.3080 & 30.4845 & 3.4 & --328 $\pm$ 129 & 5 $\pm$ 4 & -- & -- & 2 \\
    23.3695 & 30.6930 & 2.1 & --63103 $\pm$ 24185 & 431446 $\pm$ 330706 & --1.6 $\pm$ 0.2 & 5.0 $\pm$ 0.8 & 1, U77 \\
    23.3904 & 30.6674 & 1.5 & --1040 $\pm$ 228 & 237 $\pm$ 104 & -- & -- & 1  \\
    23.4085 & 30.5548 & 1.7 & --89 $\pm$ 29 & 1 $\pm$ 1 & --0.5 $\pm$ 0.2 & 1.8 $\pm$ 0.4 & 1 \\
    23.5019 & 30.6894 & 0.8 & --86 $\pm$ 28 & 6 $\pm$ 4 & --0.3 $\pm$ 0.1  & 0.7 $\pm$ 0.1 & 2 \\
    23.5067 & 30.5403 & 2.5 & --214 $\pm$ 49 & 4 $\pm$ 2 & -- & -- & 2 \\
    23.5357 & 30.6563 & 1.5 & --257 $\pm$ 114 & 14 $\pm$ 12 & -- & -- &  1\\
    23.5625 & 30.6886 & 2.0 & --253 $\pm$ 21 & 9 $\pm$ 1 & --0.1 $\pm$ 0.2 & 0.1 $\pm$ 0.1 & 1 \\   
    23.5891 & 30.6612 & 2.6 & --1352 $\pm$ 410 & 137 $\pm$ 82 & -- & -- & 2 \\ 
\hline\multicolumn{8}{c}{\bf Old, Metal-Poor Group}\\
    23.4375 & 30.7963 & 2.5 & --154 $\pm$ 9 & 2.0 $\pm$ 0.2 &  --1.3 $\pm$ 0.2  & 10.4 $\pm$ 1.2 & 1, U49 \\
    23.4672 & 30.4843 & 3.0 & --164 $\pm$ 45 & 1.0 $\pm$ 0.8 & --1.1 $\pm$ 0.2  & 9.9 $\pm$ 3.6 & 1, [SSA2010]1566 \\ 
    23.7171 & 30.4844 & 6.8 & --3129 $\pm$ 300 & 105 $\pm$ 20 & --1.1 $\pm$ 0.2 & 10.0 $\pm$ 2.0 & 2, H38 \\
    23.9250 & 28.8211 & 35.1 & --1418 $\pm$ 404 & 0.8 $\pm$ 0.5 & --1.5 $\pm$ 0.2 & -- & 2, M33-EC2 \\
    24.0088 & 29.9637 & 18.8 & --712 $\pm$ 326 & 0.7 $\pm$ 0.6 & --1.2 $\pm$ 0.2 & 10.0 $\pm$ 2.0 & 2, H33B \\
\hline
  \end{tabular}
  \label{tab:retro}
\end{table*}

\section{Star Cluster Data} \label{sec:data}
To build our catalog of stellar clusters, we combined M33 cluster spectroscopic samples available from the literature \citep{chandar+2002,sharina+2010,beasley+2015}, resulting in a total of 142 unique sources with radial velocity (RV) measurements. For 76 clusters in these studies, there are also derived ages and metallicities from integrated light spectroscopy. We find a general good agreement on the RVs between the different datasets: the RV discrepancies between \cite{chandar+2002} and \cite{beasley+2015} for 24 clusters in common yield a small mean offset of $3.0\pm5.2$ \kms. It should be noted that several cluster candidates are potential MW foreground stars and we eliminated six cluster candidates that we consider to be MW stars based on their \emph{Gaia} DR3 parallaxes and proper motions, consistent with the conclusions of \cite{Larsen2022}.

Three additional clusters are removed using a colour-colour diagram in optical photometric passbands: Figure~3 of \cite{demeulenaer+2015} shows how the ($r^{'}$ - $i^{'}$) - ($g^{'}$ - $r^{'}$) plane traces a well-defined sequence of MW stars previously misclassified as M33 clusters. Figure~\ref{fig:figure1} shows the colour-colour diagram for the clusters using the $g^{'},r^{'},i^{'}$ filters \citep{sanroman+2010}. Interestingly, the clusters removed using \emph{Gaia} astrometry as potential MW objects sit along the previously identified MW stellar sequence \citep{demeulenaer+2015}. We also found three more objects previously classified as clusters without \emph{Gaia} proper motions that trace the sequence, and we remove them from our database. Our sample contains the objects U49, R12, R14, M9, U77, H38, C20, C38, H10, U137, CB28, HM33B that are \emph{bona fide} globular clusters based on definitive colour-magnitude diagrams from resolved Hubble Space Telescope (HST) photometry \citep{sarajedini+2000,larsen+2018}\footnote{Throughout this paper, we refer to this group as the ``\emph{bona fide} clusters''.}; for those having $g^{'},r^{'},i^{'}$ photometry, they are marked with yellow circles in the left panel of Figure \ref{fig:figure1}. For nine of these GCs, ages and metallicities are available. A further sanity-check to identify Galactic foreground stars in our cluster sample is presented in the right panel of Figure~\ref{fig:figure1}. The \emph{Gaia} DR3 BP/RP excess factor is a useful parameter for finding extended sources. We use the curve from Eq. (2) by \cite{Hughes2021} to remove potential MW foreground stars. A total of five objects classified as clusters are below the curve; we identify them as non-clusters and remove them from our final list.

\begin{figure*}
\centering
\begin{subfigure}[T]{0.48\linewidth}
  \includegraphics[width=\linewidth]{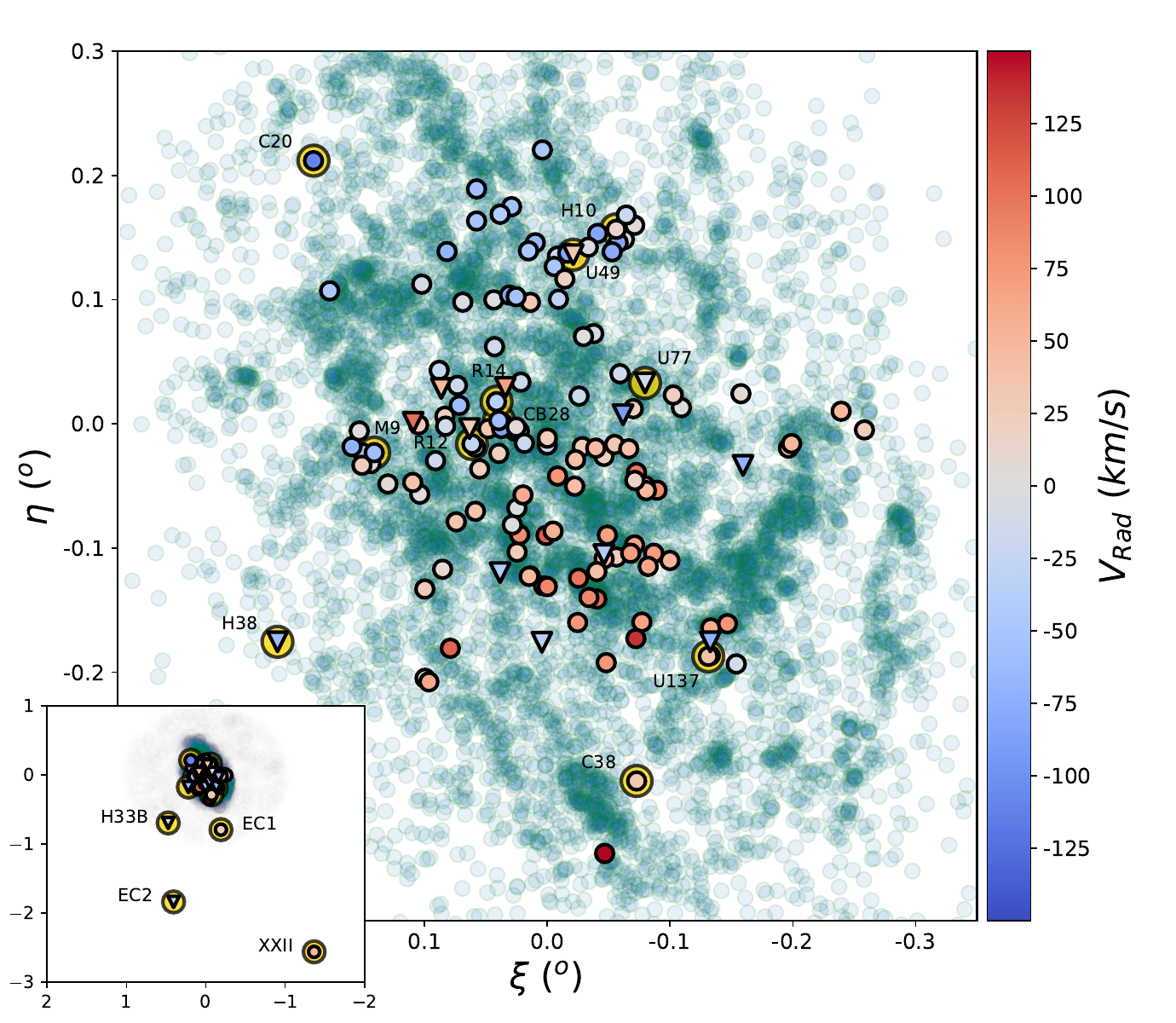}
\end{subfigure}%
\hspace{0.02\linewidth}
\begin{subfigure}[T]{0.48\linewidth}
  \includegraphics[width=\linewidth]{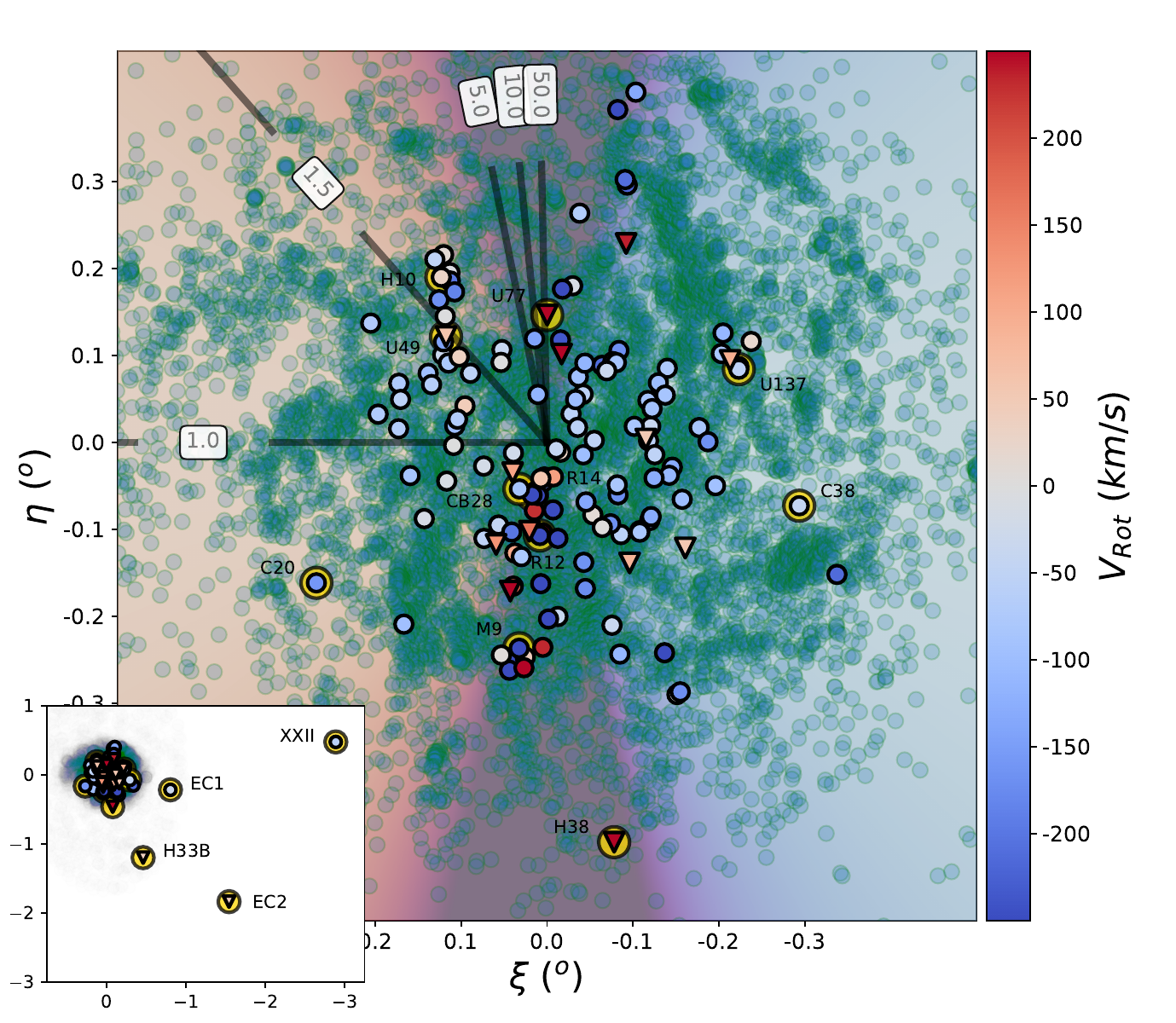}
\end{subfigure}
\caption{\emph{Left panel:} Projected spatial distribution of the M33 cluster sample colour-coded by their radial velocity corrected for the systemic radial velocity of M33. The circular markers reflect those clusters on prograde orbits, orbiting with the stellar disk, whilst the triangle represent those identified as being on retrograde orbits. The clusters lying on the large yellow circles represent the \emph{bona fide} globular clusters. In the background, we show the position of stellar objects in \emph{Gaia} DR3. \emph{Right panel:} As in the left panel, but now corrected for position angle and inclination for both the stellar systems and the \emph{Gaia} stellar objects. Here the points are colour-coded by their rotational velocity (see Section~\ref{sec:bulk}). The shaded area represents where $\cos{\theta}$ is small and any difference between $V_{\mathrm{Rad}}$ and $V_{\mathrm{sys}}$ is amplified—along with the associated uncertainties in $V_{\mathrm{Rot}}$. The black lines illustrate the $1/\cos{\theta}$ term. The smaller inset panels present the larger-scale view of the M33 system, revealing the stellar cluster systems at large distances.}
\label{fig:figure2}
\end{figure*}

The left-hand panel of Figure~\ref{fig:figure2} shows the projected spatial distribution of the cluster sample overlaid on \emph{Gaia} DR3 data centred on M33. The symbols are colour-coded by the observed radial velocity  of the clusters corrected for the systemic radial velocity of M33. The circular markers reflect the prograde population, whilst the triangles represent those clusters on retrograde orbits (see Section~\ref{sec:bulk} below for the definition of these populations). The clusters shown superposed on large yellow circles represent the \emph {bona fide} globular clusters. The insets to Figure~\ref{fig:figure2} show the projected spatial distribution of the most remote M33 clusters: H33B, EC1, EC2 and XXII. The right-hand panel of this figure shows the same population rotated to bring the major axis onto the x-axis, and then corrected for inclination. In this case, the stellar clusters have been colour-coded with their determined rotational velocity (see Section~\ref{sec:bulk} below). We have added a shaded region to better illustrate the grey scale, and included reference lines to highlight the 1/cos($\theta$) dependence. In the shaded area, where any difference between $V_{\mathrm{Rad}}$ and $V_{\mathrm{sys}}$ is amplified—along with the associated uncertainties in $V_{\mathrm{Rot}}$ as the angle $\theta$ get small. We identify four retrograde clusters, including the U77 bona-fide one previously discussed, as well as H38. This region also contains several prograde clusters, notably the metal-poor and ancient bona-fide clusters R12 and M9. Among all cases, U77 emerges as the most extreme example, as reported in Table~\ref{tab:retro}.

\begin{figure*}
\centering
\includegraphics[width=1.\linewidth]{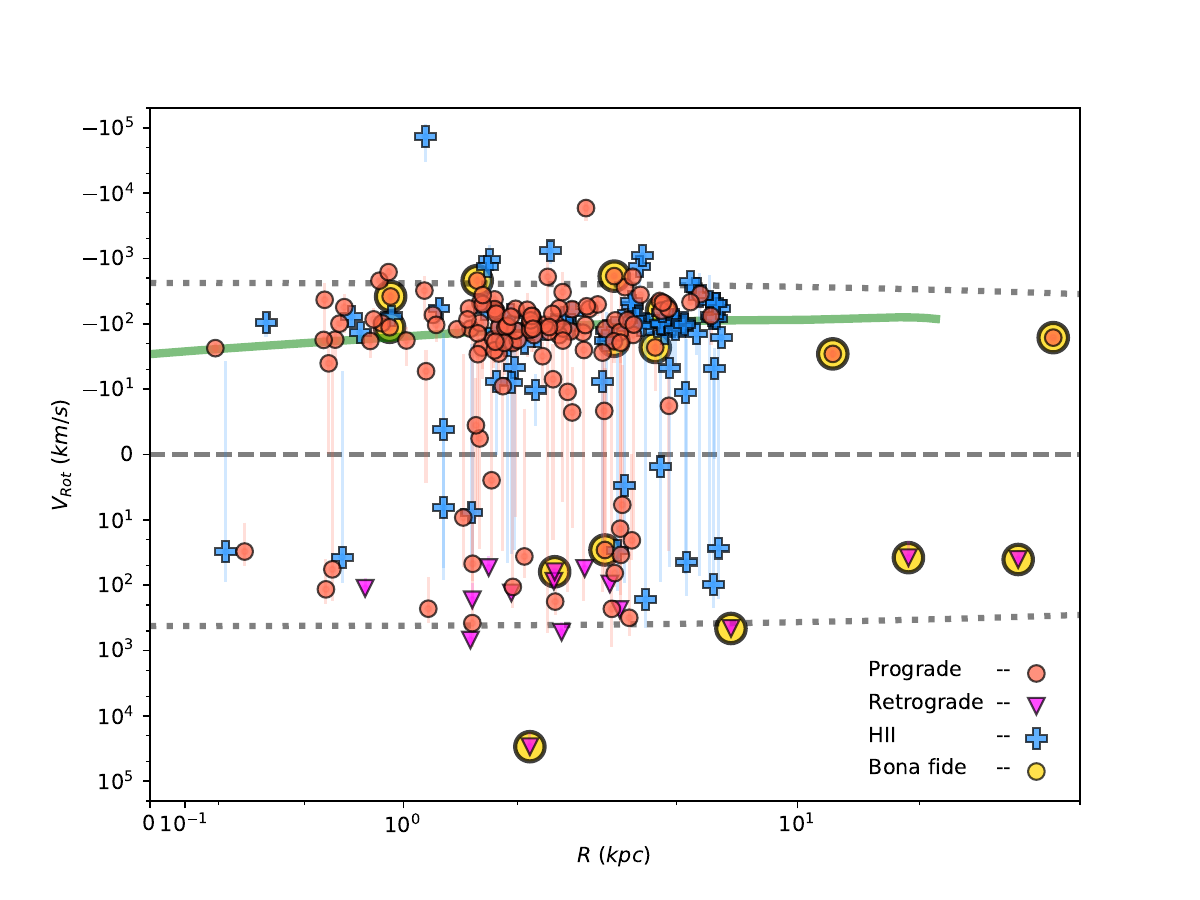}
\caption{Rotational velocity as a function of radius for the cluster population with \emph{red circles} denoting the prograde population and \emph{magenta triangles} denoting the retrograde population (see Section~\ref{sec:bulk} to see the kinematic distinction between these two population. Those stellar cluster superposed on the \emph{larger yellow circles} representing the \emph{bona fide} systems. The \emph {blue crosses} represent the \HII\ regions with the green band is the M33 rotation curve derived by \citet{2014A&A...572A..23C}, whilst the dotted lines correspond to the escape velocity of the M33 system. The uncertainties on the stellar clusters and  \HII\ regions are presented faintly as not to over-clutter the figure. As discussed in Section~\ref{sec:bulk}, whilst some \HII\ regions possess a mean retrograde velocity, none of these are statistically significant and we interpret all gas regions to be on prograde orbits.}
\label{fig:rotation}
\end{figure*}

\begin{figure*}
\centering
\includegraphics[width=1.05\linewidth]{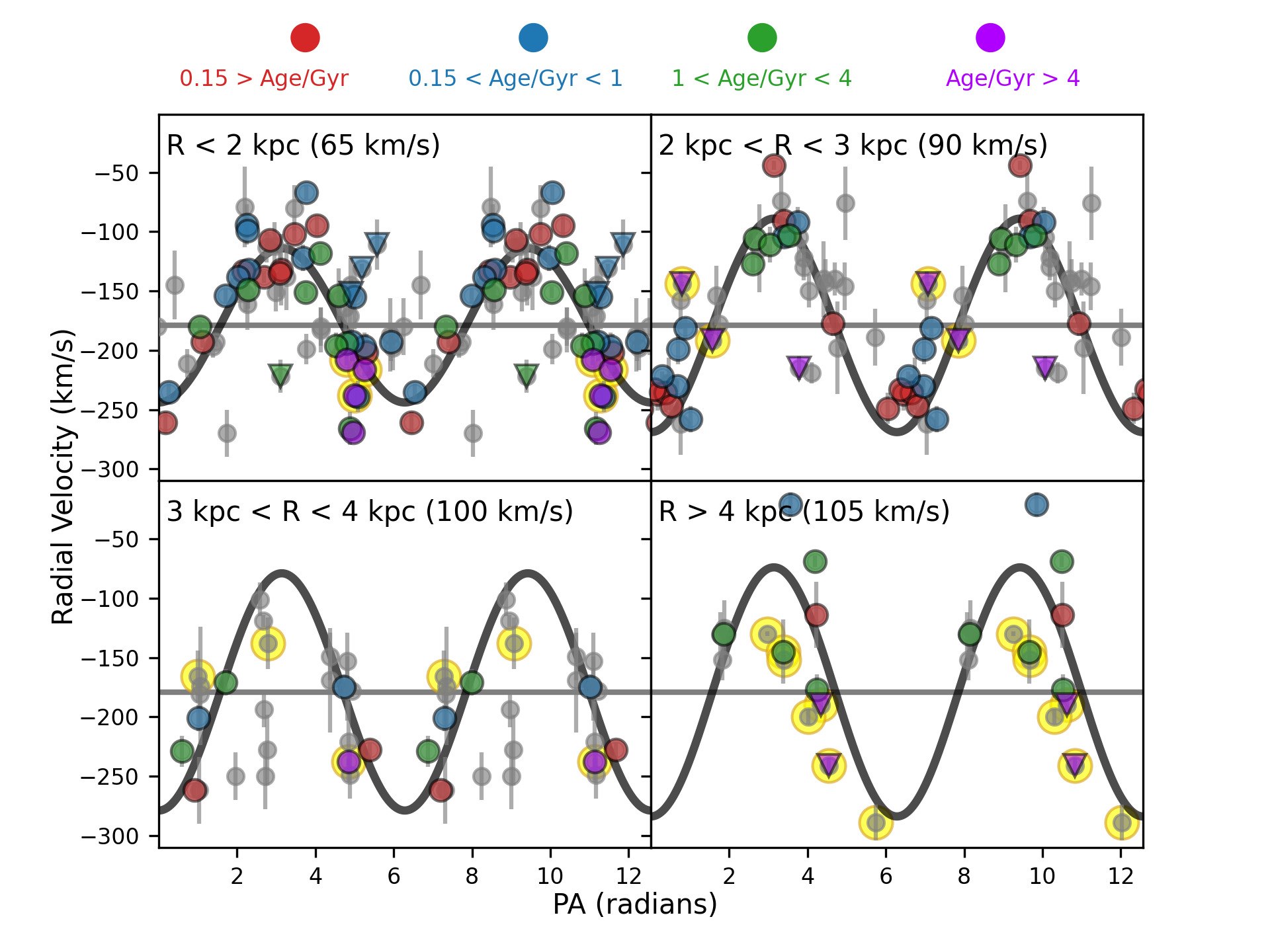}
\caption{Radial velocities as a function of position angle with respect to the major axis of M33, shown for cluster subsamples in the following radial bins: \( R > 2\,\mathrm{kpc} \), \( 2\,\mathrm{kpc} < R < 3\,\mathrm{kpc} \), \( 3\,\mathrm{kpc} < R < 4\,\mathrm{kpc} \), and \( R > 4\,\mathrm{kpc} \). The black lines represent the rotation curve at the corresponding radius, and the number in parentheses indicates the rotation amplitude. The horizontal gray line marks the systemic radial velocity of M33. When age estimates are available, the clusters are also grouped into four age bins, each represented by a different color. The triangles are denoting
the retrograde population and circles with a yellow background are the \emph{bona-fide} clusters.}
\label{fig:RV_PA}
\end{figure*}

\section{The Kinematics of the M33 Clusters} \label{sec:targets}

\subsection{Bulk Properties}
\label{sec:bulk}
With only sky positions and radial velocities, we have limited spatial and kinematical information on the properties of M33's star clusters. Here we make the assumption that the clusters lie in the disk of M33 such that we can deproject their properties; we discuss the implications of this assumption in Section~\ref{subsection:caveats}. 

\begin{figure}
\centering
\includegraphics[width=1.05\linewidth]{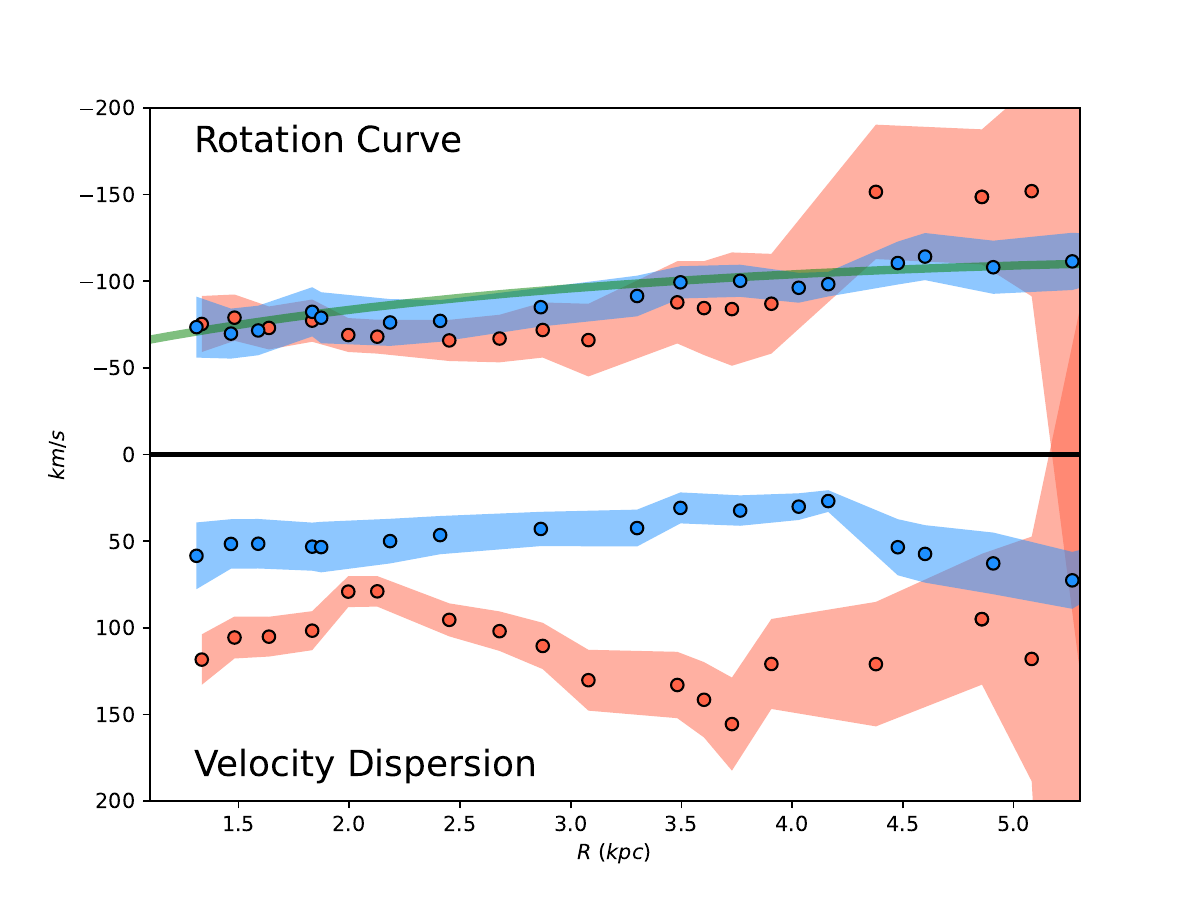}
\caption{The rotation velocity and velocity dispersion of the cluster population in red and the \HII\ regions in blue. The points correspond to the best fit value of the quantities, with the coloured bars denoting $1\sigma$ uncertainties in these quantities. The underlying green denotes the rotation curve of M33, with both populations following this rotation. Clearly, the cluster velocity dispersions far exceed those of the \HII\ regions.}
\label{fig:dispersion}
\end{figure}

With this assumption, we can determine the rotational velocity, $V_{Rot}$ of different points across the face of a disk galaxy as $V_{Rad} = V_{0} + V_{Rot}\sin{\xi}\cos
{\theta}$, where $V_{Rad}$ is the measures radial velocity of the stellar cluster,
$V_{0}$ is the systemic bulk radial velocity of the system \citep[in the case of M33, $V_{0}$ is measured to be $-179\pm1$\kms;][]{Koch2018}, $\xi$ is the inclination angle between the line of sight and the normal to the plane of the galaxy, $V_{Rot}$ is the rotational velocity within the plane at a radial distance $R$, and $\cos{\theta} = X/R$, where $X$ is the position along the major axis \citep{Rubin_Ford1970}. 
We calculate the parameter $X/R$ for each cluster, assuming a position angle of the major axis of 22.5$^{\circ}$ and an inclination angle of 54$^{\circ}$. Furthermore, a distance to M33 of 840\kpc\ is assumed. The uncertainties of the radial velocity are propagated into the rotational velocity by Monte Carlo resampling. 

Figure~\ref{fig:rotation} shows M33 rotational velocities as a function of radius. The cluster population is represented by circles and triangles, with the triangles representing the retrograde population (for a definition, see below). To understand the motion of the galactic disk, we also consider the motion of gas (blue crosses), using the radial velocities for seventy \HII\ regions in M33 determined in LAMOST DR7 \citep{Alexeeva2022}. Around 30$\%$ of the measurements show uncertainties in radial velocity larger than 20\kms.%


It is important to note that whilst there are \HII\ sources with apparently positive velocities in this figure such that they appear on retrograde orbits, these velocities are not statistically significant as $v-2\sigma$ for all of the \HII\ sources are negative; with 70 sources, there is a 20\% chance that none will be seen as $>2\sigma$ (one-sided) and so this result is consistent with all of the \HII\ sources being on prograde orbits. The green band is the M33 rotation curve, as derived by \citet{Corbelli2000} using the hydrogen 21-cm line, whilst the dotted lines correspond to the escape velocity of M33 (see Sect.~\ref{sec:escape_velo}). The uncertainties of the velocity are presented in the figure, but are kept relatively faint as not to overly clutter the figure.


To compute the mean and dispersion of the gas and cluster samples, we fit them to Gaussians in bands 2\kpc\ thick at 1\kpc\ intervals. The fits and uncertainties of the rotational velocity and the velocity dispersion are presented in Figure~\ref{fig:dispersion}, with the clusters presented in red and the \HII\ regions in blue. Again, the green represents the rotation curve of M33 and it is clear that, like the \HII, the stellar clusters predominately rotate with disk like kinematics. We note that the \HII\ is dynamically cold, with a velocity dispersion of $\sim50$\kms, as expected for this population, whereas the inferred velocity dispersion for the cluster population is hotter, with $\sim130$\kms. The kinematical character of the cluster population, in particular its low and decreasing $V_{Rot}/\sigma_{Rot}$ with radius, suggests that some of clusters in our sample, particularly those at large $R$, are potentially associated with the halo of M33 (see also \citealt{schommer+1991}). Finally, in Figure~\ref{fig:RV_PA} we have the radial velocities as a function of position angle with respect to the major axis of M33. The figure is divided in four radial bins:  \( R > 2\,\mathrm{kpc} \), \( 2\,\mathrm{kpc} < R < 3\,\mathrm{kpc} \), \( 3\,\mathrm{kpc} < R < 4\,\mathrm{kpc} \), and \( R > 4\,\mathrm{kpc} \), and color-coded by age when available. We use the rotation curve model describe above at the corresponding radius and show it as black lines in the radial velocity–position angle plane. The number in parentheses indicates the rotation amplitude. We find that most clusters follow the disk rotation, while some retrograde clusters clearly stand out, represented as triangles in the figure. There is no obvious phase difference between older and younger clusters as noted in \cite{beasley+2015}. However, we note a group of bona-fide clusters at \( R > 4\,\mathrm{kpc} \) that may show a small phase shift of \(-0.5\) to \(-1\) radians.

\subsection{Asymmetric drift correction}

\cite{beasley+2015} corrected the rotational velocity of their clusters for asymmetric drift, the difference between the velocity of a hypothetical set of stars possessing perfectly circular orbits and the mean rotational velocity of the population under consideration. This is a phenomenon observed in the Milky Way disk, where old and metal-poor stars are lagging behind with respect to younger disk populations \citep[e.g.,][]{Golubov2013}. Although the asymmetric drift in M33 is not well understood and this correction is more suitable for clusters with disk-like kinematics rather than for a halo stellar population, we have assessed its effect in our kinematical analysis.

To estimate the velocity dispersion after subtracting the rotation curve, we have binned the data based on radial distance $R$. 
The values we obtain are: $\sigma_{\rm los}$ = 40.3 km/s for $R$ $<$ 2 kpc, 
$\sigma_{\rm los}$ = 37.1 km/s for 2 $<$ $R$ $<$ 3 kpc, $\sigma_{\rm los}$ = 44.1 km/s for 3 $<$ $R$ $<$ 4 kpc, and $\sigma_{\rm los}$ = 53.3 km/s for $R$ $>$ 4 kpc. We estimate that $\sigma_{\rm los}$ is about 40 -- 50 km/s for the cluster sample. We applied the asymmetric drift correction using the estimated velocity dispersion. For this, we employed equation (2) from \cite{beasley+2015}. We initially assumed a velocity dispersion of zero, which should have no effect on the correction. Then, we tested values of 20 km/s and 45 km/s. The correction is applied as a squared term, however its overall influence remains minimal, even in the case of the retrograde sample. This suggests that the asymmetric drift correction does not significantly alter the overall kinematical trends observed in our dataset. 

\subsection{The Retrograde Cluster Population}
\begin{figure}
\centering
\includegraphics[width=\linewidth]{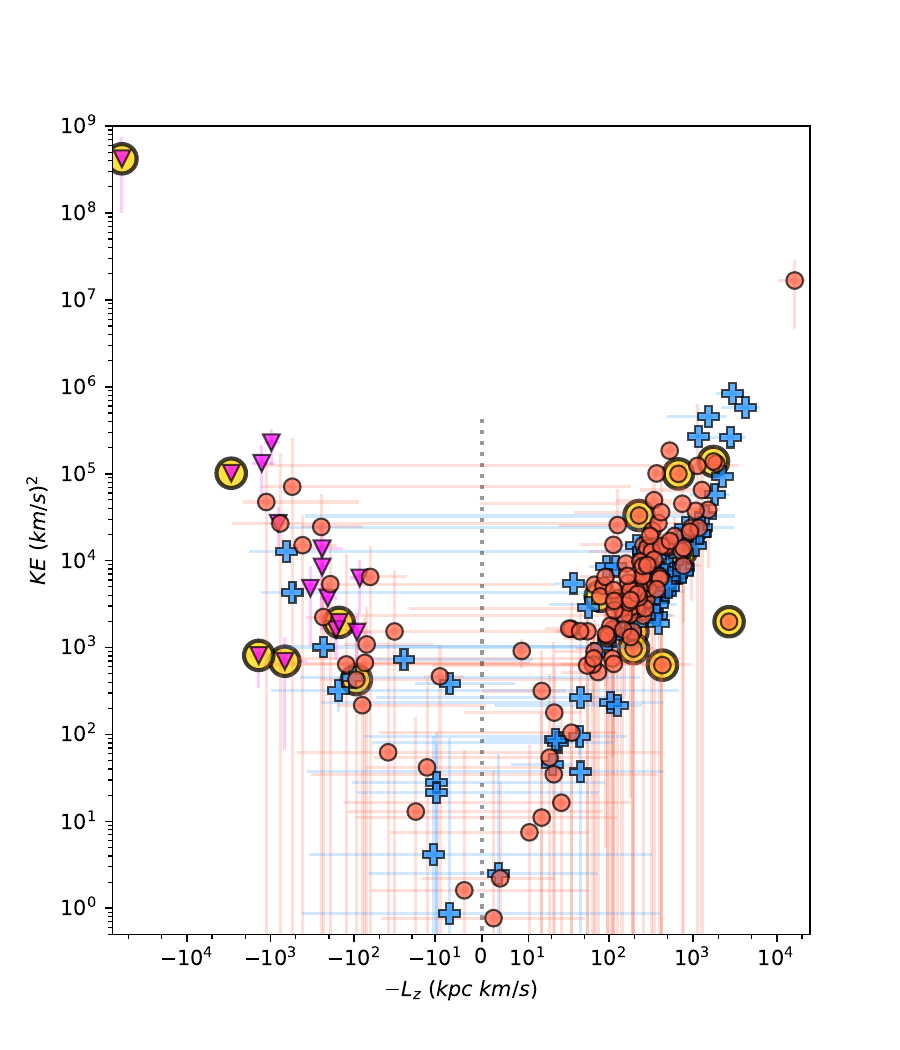}
\caption{The kinetic energy as a function of the vertical component of the angular momentum for the M33 \HII\ regions (\emph{blue cross symbols}), and the M33 cluster system (\emph{solid circles} and \emph{triangles}). As expected, the \HII\ regions traces the prograde motion of the galaxy. Cluster candidates in retrograde motion are labelled by \emph {triangles}. The objects shown with a symbol superposed on a \emph{large yellow circle} are the set of \emph{bona fide} clusters. Note that some of the error bars are smaller than the plotted symbols. Note also the scaling of the axes (\emph{asinh}) to accommodate all of the data within one figure.}
\label{fig:gas_stream}
\end{figure}

\begin{figure}
\centering
\includegraphics[width=\linewidth]{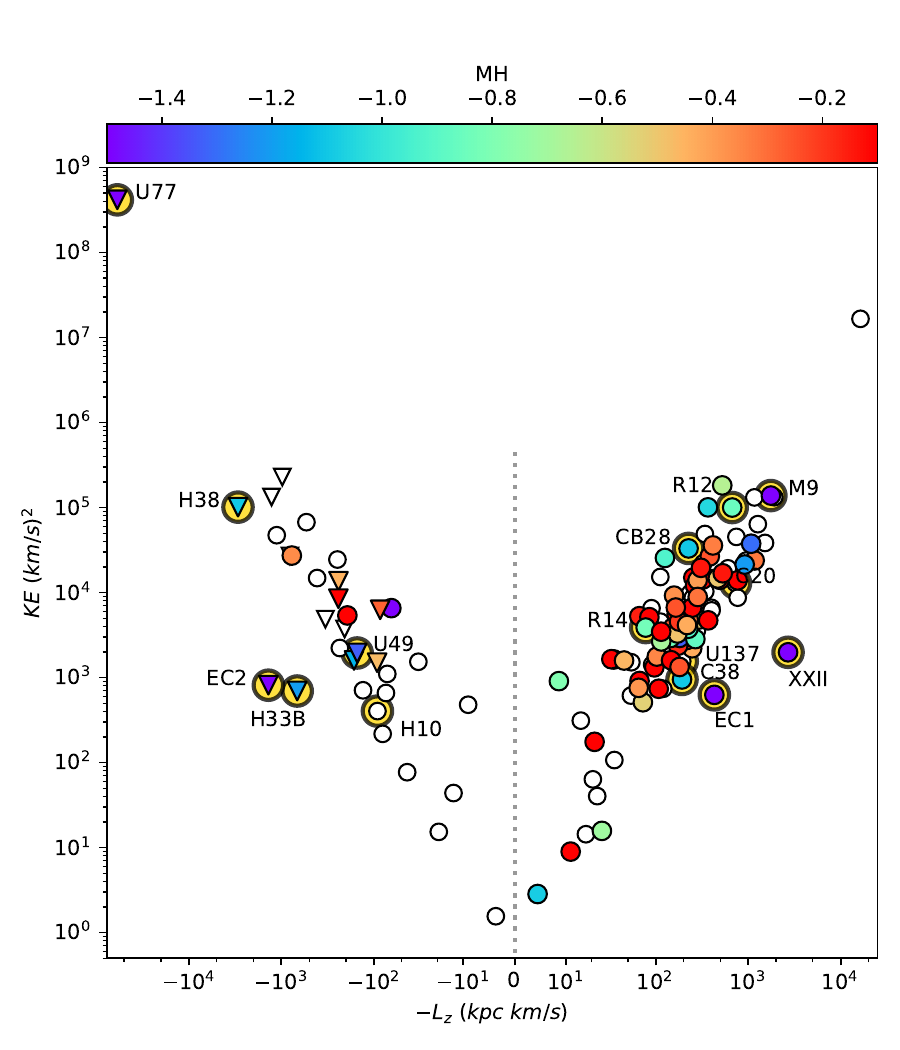}
\includegraphics[width=\linewidth]{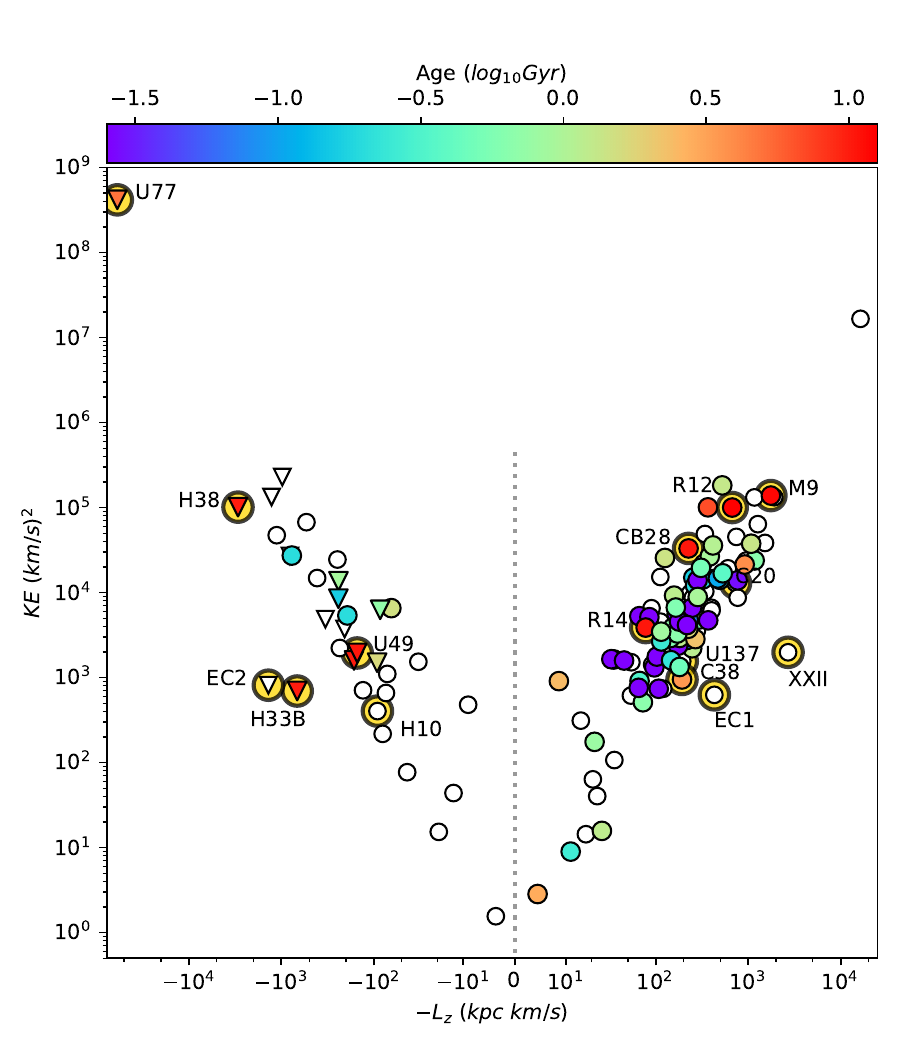}
\caption{The same as Fig.~\ref{fig:gas_stream} but with the points colour-coded by metallicity (\emph{upper panel}) and $\log_{10}(ages)$ (\emph{bottom panel}). Clusters tracing the rotation of the galaxy are predominantly young and metal-rich. Symbol shapes are the same as in Fig.~\ref{fig:gas_stream}. Note the scaling of the axes (\emph{asinh}) to accommodate all of the data within one figure.}
\label{fig:stream}
\end{figure}

Beyond measuring the rotational velocity for the cluster population, it is possible to estimate the angular momentum in the vertical component $L_{z} = -V_{Rot} \cdot R$; note that the minus sign is to ensure that prograde orbits have a positive angular momentum, whilst retrograde motions are negative. Using this quantity together with the kinetic energy ($KE \sim V_{Rot}^{2}/2$), we build \emph{pseudo-Lindblad diagrams} for the cluster population and the \HII\ regions, as shown in Figures~\ref{fig:gas_stream} and \ref{fig:stream}. We define members of the retrograde population as those where $-Lz + 2\sigma_{Lz} < 0$ to remove those clusters whose negative velocity is not statistically significant. In the figures shown hereafter, the most certain retrograde clusters defined in this robust manner are represented by triangles. There are 15 star clusters out of a total sample of 145 that meet the criteria for being on retrograde orbits. To judge the statistical significance of this number, the probability of seeing more than ten sources at $>2\sigma$ (one-sided) out of a sample of 145 is 0.06\%, and hence we conclude that this population of retrograde clusters is significant. As noted previously, the \HII\ regions in Figure~\ref{fig:gas_stream} clearly traces the prograde motion of the disk (blue cross symbols). The vast majority of the M33 cluster population is seen to be in prograde motion (solid orange circles in Fig.~\ref{fig:gas_stream}).

However, Figure~\ref{fig:gas_stream} also shows the clusters that are most probably in retrograde motion (marked as triangles). 
Figure~\ref{fig:stream} shows figures similar to those discussed above, but now colour-coded by metallicity, $[M/H]$, and $\log{(age)}$, when these quantities are available. Most of the prograde clusters trace the young and metal-rich disk of M33. There are 18 old star clusters with lower age limits of $\log_{10}(t/\gyr) \sim 9.0$ in our sample. Furthermore, eight of these clusters are very ancient ($\log_{10}(t/\gyr) > 10.0$), and seven of them are among the \emph{bona fide} clusters (CB28, M9, U49, H33B, R12, R14, H38).

In Table~\ref{tab:retro}, we list the 15 clusters of our original sample of retrograde clusters that satisfy $-L_z + 2\sigma_{L_z} < 0$. Five of them are \emph{bona fide} globular clusters (U77, U49, H38, M33-EC2, and H33B). Interestingly, four clusters — U49, SSA2010, H38, and H33B — form a distinct group of old (approximately 10 Gyr), metal-poor clusters. Another cluster in retrograde motion, M33-EC2, is also metal-poor ([M/H] $\sim$ --1.6) and, although there is no formal age estimation for this object, according to \cite{huxor+2009} it has colours consistent with those of an old to intermediate-age globular cluster. Thus, we group M33-EC2 with the other four metal-poor clusters as part of an "Old, Metal-Poor Group" that may share a common origin, perhaps as part of an early accretion event. It is remarkable that the retrograde clusters with the largest projected separation, $R$, are part of this Old, Metal-Poor Group.

Meanwhile, object U77, which is clearly in retrograde motion with respect to the HI gas in our analysis and has a very large kinetic energy (KE / 10$^{3}$ $\sim$ 4 $\times 10^{5}$\KE), is one of the most metal-poor clusters ([M/H] = --1.6 $\pm$ 0.2) in the M33 galaxy, yet it has an intermediate age of $5\pm1$\gyr\ \citep{beasley+2015}. The discovery of a genuine intermediate age star cluster in M33 is not new \citep{chandar+2006} --- for example, the cluster C38 with [M/H] = $-1.1\pm0.3$ has an age of $3.5\pm1.5$\gyr. Interestingly, C38 has a slightly prograde motion with a large uncertainty associated with the angular momentum in the vertical component ($-L_z = 190 \pm 150$ \kpckms) and a low kinetic energy (KE / 10$^{3}$ = 0.9 $\pm$ 0.6 \KE). This cluster could also be part of the non-rotating M33 halo component discussed above. U77's 5 \gyr\ age might then represent the upper limit of a relatively recent merger that formed this component of M33's halo population; however, as discussed in Section \ref{sec:escape_velo}, it is likely that this object is not bound to M33.

Table~\ref{tab:retro} also shows the equatorial coordinates, radial distance, vertical component of the angular momentum, kinetic energy, and metallicity and ages, when available, for these particular clusters. In the Notes column, we show (a) the names of the {\emph{bona fide} clusters, (b) the number 1 if there is Str\"omgren photometry and \emph{Gaia} BP/RP excess factor values for these clusters, and/or (c) the number 2 if there are \emph{Gaia} BP/RP excess factor values and these objects are labeled as extended sources (see Sect.~\ref{sec:data}), but there is no Str\"omgren photometry for these objects. As mentioned in Section~\ref{sec:data}, the ($r^{'}$ - $i^{'}$) - ($g^{'}$ - $r^{'}$) plane can trace MW stellar objects compared to extended objects in M33.

Not listed in Table~\ref{tab:retro} is a {\it bona fide} cluster, H10, with a large uncertainty in angular momentum, i.e., $-L_z = -104$ $\pm$ 177 \kpckms; this cluster also has low kinetic energy, (KE / 10$^{3}$) = 0.4 \KE. The estimated metallicity of H10 obtained by isochrone fitting to the cluster's colour-magnitude diagram (CMD) sequence is [M/H] = --1.4 $\pm$ 0.3 \citep{sarajedini+1998}. There is a similar {\it bona fide} cluster, U49, that has $-L_z = -161\pm9$ \kpckms and low kinetic energy, (KE / 10$^{3}$) = 2 \KE. This cluster has a similar metallicity to H10 and is very old ($\sim$ 10\gyr). Table~\ref{tab:retro} also shows an old cluster ([SSA2010]1566, at age 9.9 $\pm$ 3.6 Gyrs) with similar kinematics. In addition, the {\it bona fide} old  ($\sim$ 10 Gyrs) cluster R14 with $-L_z = 83\pm 13$\kpckms\ and low kinetic energy (KE / 10$^{3}$) = $4 \pm 1$ \KE) could also be part of this group. Although not well established, M33 could have a weak pseudobulge with a weak bar structure of 1\kpc\ length \citep{Corbelli2007}. If so, this could lead to potential misclassifications between the disk and bulge populations. These four ancient and metal-poor clusters could be part of a pristine, non-rotating halo component of M33 \citep{chandar+2002,mcconnachie2006,Cullinane2023}, however, their low kinetic energy might indicate they are part of the pseudobulge population. 

\begin{figure*}
\centering
\includegraphics[width=1.\linewidth]{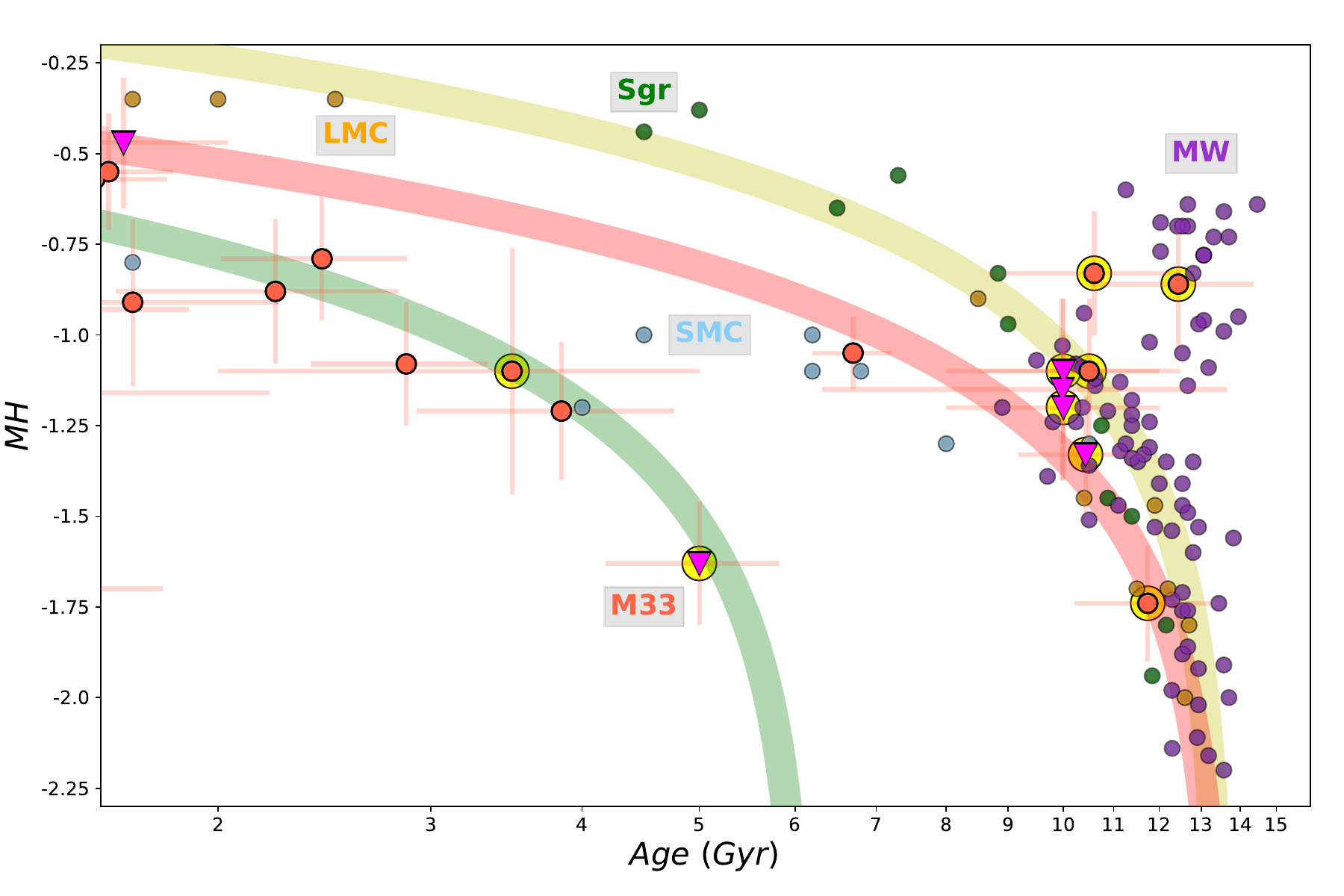}
\caption{The age-metallicity relation of the M33 cluster population. The figure also shows the star clusters in the MW (purple), Sgr dwarf spheroidal galaxy (green), the LMC (dark yellow circles), and SMC (light blue). As with the other figures, the \emph{magenta triangles} indicate retrograde motion according to our analysis in this study, whilst the \emph{orange circles} are the prograde clusters. Eight of these clusters are also \emph{bona fide} clusters (indicated with the  background \emph{yellow circle}). The three different lines represent a simple closed-box chemical evolution model assuming a constant star formation rate (see text for details). Note also that the late chemical enrichment is relevant to understanding the origin of U77 and that this cluster may be significantly older than the age inferred in \protect\cite{beasley+2015}}.
\label{fig:AMR}
\end{figure*}

In our sample (see Figure~\ref{fig:gas_stream}), we also have four other \emph {bona fide} clusters with high angular momentum and low kinetic energy. One is H33B, which has a large radial distance, $R \sim 18$\kpc, and is in retrograde motion relative to the M33 HI gas. This ancient and metal-poor cluster (see Table~\ref{tab:retro}) was first reported to be remote, compact, and with high surface brightness \citep{huxor+2009}. Furthermore, M33B has $\alpha$-element abundances closer to scaled solar values and significantly different abundance patterns than the other M33 clusters analysed using high-resolution spectra, suggesting an accretion origin for M33B \citep{Larsen2022}. Another cluster of this kind, also in retrograde motion, is M33-EC2. This system is metal-poor ([M/H] $\sim$ --1.6) and, according to \cite{huxor+2009}, EC2 has colours consistent with those of an old to intermediate-age globular cluster. A cluster in our analysis with similar age and metallicity characteristics but with prograde motion is M33-EC1 \citep{huxor+2009}, and this is also 
the case with the potential dwarf spheroidal galaxy M33-XXII \citep{chapman+2013}. These two objects have relatively low rotational velocity but a large radial distance.

Finally, we discuss our findings in light of the cluster orbits in the E-MOSAICS simulations \citep{Pfeffer2020}. Using these cosmological, hydrodynamical simulations of galaxy formation, \cite{Pfeffer2020} found that accreted clusters are sensitive to the mass of the satellite galaxy, where earlier mergers and larger galaxy masses deposit clusters at lower total orbital energy. The simulation results suggest that the low kinetic energy and ancient clusters mentioned above may have been part of a major earlier merger in the history of M33. Interestingly, \cite{Pfeffer2020} also found that \emph{in situ} clusters could have retrograde orbits, but generally, their reported orbits tend to have lower energies than the accreted clusters in their simulations. To better understand the origin of the retrograde clusters in M33, we study the age-metallicity relation in the next section.

\subsubsection{The age-metallicity relation}
The relation between age and metallicity, known as the age-metallicity relation (AMR), is a helpful construct to constrain the evolutionary history of a galaxy via the chemical enrichment of its individual stellar and cluster components \citep{Anguiano2012,Kruijssen2019}. Moreover, we can use the AMR to estimate the fraction of \emph{in situ} formed star clusters versus accreted ones \citep{forbes+2010,massari+2019,Woody2021}.

Figure~\ref{fig:AMR} shows the ages and metallicities of M33 clusters older than 1 Gyr. The figure also shows the ages and metallicities for old star clusters in the MW \citep[][purple circles]{Kruijssen2019}, Sgr dwarf spheroidal galaxy \citep[][green circles]{forbes+2010}, LMC (dark yellow circles), and SMC (light blue circles) (from the appendix in \cite{Horta2021} using a confidence code of 1). The triangles indicate retrograde motion according to our analysis in this study. Furthermore, seven of these clusters are very ancient, older than 10\gyr, and are also \emph {bona fide} clusters (CB28, M9, U49, H33B, R12, R14, H38). The age-metallicity relation for M33 reveals three potential sequences. Two of them resemble the two distinct tracks observed in the MW \citep[e.g.,][]{Marin_Franch2009,forbes+2010}, one with a very old population, and the other branch with a range of ages and metallicities, containing clusters from the accreted Sgr dwarf galaxy \citep{Law2010,Bellazini2020}. In the M33 cluster sample, we also observe a third track containing clusters younger than 6\gyr\ and with a range of metallicity from $-1.8 < [M/H] < -0.7$. One of these clusters (U77, see discussion about this cluster in Sect.~\ref{sec:escape_velo}) is in retrograde orbital motion. We can describe the observed age-metallicity relation with a simple closed-box chemical evolution model assuming a constant star formation rate.\footnote{For a similar approach using MW clusters see \cite{forbes+2010,massari+2019}.} The green dashed line represents the model following the accreted clusters from the Sgr galaxy. The red line is a close-box model with a continuous star formation rate to represent one of the tracks in the M33 AMR, and finally, the green line is the model for the youngest track in the AMR. The oldest clusters with a metallicity range of $-1.8< [M/H] < -0.7$, are probably formed \emph{in situ}, while the population of clusters following the model represented by the red line have clusters on retrograde orbits (\emph{magenta triangles}), and they are candidates for having been accreted from a satellite galaxy. Similarly, we have the cluster population lying along the chemical model represented by the green line.  This population might have been involved in a more recent and less massive accretion event. However, we need to bear in mind that the estimated ages and metallicities from integrated light spectra might suffer from age-metallicity degeneracy \citep{beasley+2015}. Also, we stress that the curves in Figure~\ref{fig:AMR} do not show fits to the data points, but rather represent how a simple close-box chemical evolution model can describe the observed AMR.

\subsection{The Escape Velocity of M33}
\label{sec:escape_velo}
The escape velocity of M33 can be calculated using the gravitational potential following $\upsilon_{esc}(r) \equiv \pm \sqrt{2|\phi(r)|}$, where $\phi(r)$ is the approximation of the gravitational potential of our M33 model. We represent the gravitational potential by a three-component model comprised of a dark matter halo following a Navarro-Frenk-White profile \citep{NFW1996} with a total mass of $3.7 \times 10^{11}$\msun\ and a scale-length of 20\kpc. The stellar and gas disks are represented as a Miyamoto-Nagai potential with a total mass of $8 \times 10^{9}$\msun, scale-length and height of 3 and 0.5\kpc, approximating the rotation curve presented in \citet{2014A&A...572A..23C}.

Figure~\ref{fig:rotation} shows the rotational velocity as a function of radius for the M33 cluster population. The figure also shows the $\upsilon_{esc}$ curves in dotted lines. The orange circles are clusters marked as prograde, while the triangles are retrograde systems. The blue crosses represent the HI gas. Almost all of the cluster population appears to be bound to M33; however, there are clearly two star systems, one prograde and one retrograde, with a velocity greater than the escape velocity. One is the \emph {bona fide} cluster U77 \citep{christian+1982,sarajedini+1998}. This cluster has an extended blue horizontal branch (BHB), contrary to most M33 halo clusters, where the HB are significantly redder \citep{sarajedini+1998}. Moreover, hot stars in the BHB can affect the spectral energy distributions and therefore the spectral age estimation for this object, making the cluster look younger \citep{beasley+2015}. Therefore, U77 may well be much older. According to our results, U77 is an unbound M33 system. We hypothesise that this object is a dwarf spheroidal galaxy within the Local Group. However, U77 shows a broad distribution for the velocity and a significant uncertainty due to its small coordinate value along the major axis (see Table~\ref{tab:retro}). Finally, the \emph{bona fide} cluster H38 also has a velocity slightly larger than $\upsilon_{esc}$ (see Fig.~\ref{fig:rotation}). This too is an ancient metal-poor object (see Table~\ref{tab:retro}).

Figure~\ref{fig:rotation} also shows a prograde cluster with a velocity much larger than the escape velocity of the M33 galaxy. The SIMBAD \citep{Wenger2000} name for this object is [SBG2007] 23, and it first appeared in the catalogue of \cite{sarajedini+2007}. The rotational velocity for this cluster is less extreme than that of U77. Unfortunately, there is no colour-magnitude diagram for this source as there is for U77. We suggest that [SBG2007] 23 is also a candidate dwarf spheroidal galaxy. 

Finally, we identify two \emph{bona fide} clusters with a prograde velocity slightly larger than $\upsilon_{esc}$ but within the errors: cluster R12 ([M/H] $= -0.9 \pm 0.2$, age $= 12.4 \pm 1.9$\gyr) and M9 ([M/H] $= -1.7 \pm 0.2$, age $= 11.7 \pm 1.5$\gyr). The clusters H33B, M33-EC1, M33-EC2, and the potential dwarf spheroidal galaxy M33-XXII \citep{chapman+2013} appear clearly in Figure~\ref{fig:rotation} at large radii ($R > 10$\kpc). All of these objects have metallicity [Fe/H] $\sim$ --1.6, and according to \cite{huxor+2009}, EC1 and EC2 have colours consistent with old to intermediate-age globular clusters. We found the four objects to be bound to M33, in agreement with the results from \cite{chapman+2013}. Two objects, H33B and EC2, are potentially in retrograde motion with respect to the rotation of the M33 disk.

\subsection{Caveats}\label{subsection:caveats}
The starting assumption underlying our analysis is that the star clusters we identify are located in the disk of M33 and are moving with disk-like motion. We note that this is borne out in Figure~\ref{fig:dispersion} which shows the star clusters (and \HII\ regions) follow the rotation curve of disk, with the cluster population showing a hotter dispersion than the gas. However, there is potential that some of the stellar systems identified in this present study are not part of the disk population but are interlopers from an M33 halo cluster population. Classical halo clusters would be kinematically distinct from the M33 disk rotation, presumably being on random, pressure-supported orbits. In the analysis presented in this paper, halo interlopers will contribute a mix of apparently prograde and retrograde clusters. However, high-redshift galaxies often show irregular, "clumpy" morphologies and disordered kinematics, where their star formation is bursty, with large fluctuations around a mean rate, driven by dynamical processes like mergers or gas infall ~\citep[e.g.,][]{Hopkins2023}.

Estimating the potential contamination of our cluster sample from a classical halo population is difficult as there is no consensus on the total globular cluster population of M33. The situation is further complicated by the fact that the M33 stellar halo appears to be very tenuous, or even non-existent ~\citep{mcmonigal+2016,Ogami2024}, potentially indicating that this component was either never formed or was stripped away in a previous strong interaction ~\citep{mcconnachie+2009}. Such an interaction would have similarly stripped, or disrupted, the halo globular cluster population and it would be difficult to reconcile a substantial population of halo globular clusters appearing in the inner regions after multiple dynamical timescales since the interaction. 
However, if we suppose that the 15 retrograde clusters are interlopers from the halo, we can expect a similar number that contribute to those displaying prograde motion, suggesting an interloper fraction of $\sim 30/145 = 21\%$ within the sample. Given that the distribution of halo globular clusters is relatively flat, this number suggests that there should be a substantial number of globular clusters throughout the halo of M33 (of order 100-200 within $\sim30$\kpc), a situation at odds with panoramic searches ~\citep[e.g.,][]{Huxor2011}.

Given the available data, comprised of the sky positions and radial velocities of the known candidate M33 star clusters, it is not possible unambiguously to identify whether the observed stellar clusters are purely a disk population, or whether it is contaminated by interlopers from the halo. Hence it will be essential to obtain further data, in particular accurate distances and proper motions, observations presently only accessible with JWST, to answer the question of whether this truly represents an accreted population into M33.

\section{Conclusions}\label{sec:conclusions}
We explored the kinematics, metallicities, and ages of the individual star clusters of M33 to infer the accretion history of the galaxy. In particular, for kinematics, we compute the vertical component of the angular momentum to find and characterise clusters in retrograde motion, which is a useful discriminant of an accreted cluster \citep{Rodgers1984,Zinn1993}. We find a group of fifteen clusters (Table~\ref{tab:retro}) with negative values in their vertical component of the angular momentum, indicating a retrograde motion. With a significant representation of clusters potentially originating through accretion, M33's cluster system seems to hold ample evidence for past merger activity. Furthermore, the ages of the retrograde star clusters in M33's halo span a wide range, which, under the accretion hypothesis, would require (1) progenitor systems featuring components with extended star formation histories, and (2) accretion events until even relatively recent times. We also stress that we have assumed that most M33 clusters are moving with disk-like motion; however, some of these objects could potentially be interlopers from an M33 halo cluster population (see Section~\ref{subsection:caveats} for more details).

M33 is surrounded by a stellar structure that has a "S-shaped" appearance, which has been found \citep{mcconnachie+2009,mcconnachie+2010} to align with the orientation of the HI disk warp. This coincidence is explained by associating these substructures with the tidal disruption of M33 in its orbit around M31. However, thanks to \emph{Gaia} proper motions, the 3D velocities of M31 and M33 are precisely known \citep{vanderMarel2019}, and these robust velocity measures suggest that M33 may be on its first infall into M31. If M33 did not have a previous close tidal interaction with M31, alternative explanations for the gas and stellar disk warps of M33 are required. For example, the accretion of a relatively massive dwarf galaxy could induce such warping. Our results show that a group of clusters observed in the halo of M33 are likely the result of merger activity, and this, together with the inference that M33 is on its first infall, may also have been responsible for creating its gaseous and stellar disk warps.

The kinetic energy of the retrograde M33 clusters spans a wide range, much wider than that of M33's \HII\ regions (Fig~\ref{fig:gas_stream}). There are also high $KE$ prograde clusters that may represent accreted M33 objects, although this suggestion is not as strongly motivated as for the retrograde objects. Among the high $KE$ sources is the object U77, which has been classified as a \emph{bona fide} cluster \citep{sarajedini+1998}, and has a very large retrograde vertical angular momentum value. This "cluster" also shows a rotational velocity greater than M33's escape velocity, $\upsilon_{esc} > 450$\kms, suggesting it is unbound, and a colour-magnitude diagram showing an extended horizontal-branch, unlike the majority of M33 halo clusters \citep{sarajedini+1998}. This object is a potentially newly identify dSph galaxy, another candidate "satellite of a satellite" for M33 \citep{chapman+2013,collins2024}. It is also remarkable that M33-XXII, a potential dwarf spheroidal galaxy \citep{chapman+2013}, as well as remote extended objects such as H33B, M33-EC1, and M33-EC2 \citep{huxor+2009}, appear to have a very wide range of angular momentum values but a very similarly low (KE / 10$^{3}$) $\sim$ 0.9 \KE. We have found that the latter four ancient and metal-poor objects are bound to M33. According to $\Lambda$CDM simulations of M33 \citep{Patel2018}, satellites remain bound to the host galaxy during its first infall into M31, and the discovery of more than four dwarf galaxies as present satellites of M33 is strong evidence for the first infall scenario for M33, and more than six are definitive.

The linear correlation between the total number of GCs in a galaxy and its halo virial mass \citep{Blakeslee1997,Spitler2009,Burkert2020} can be used to estimate a lower limit for the accreted mass in the halo of M33. We obtain a lower limit of $M_{\rm vir} \sim (7 \pm 3) \times 10^{10}$\msun.\footnote{To compute the uncertainties in the virial mass we have taken into account clusters that appear to have error bars that allow them to be in a prograde or retrograde orbit by taking the $\sqrt{N}$ clusters in the correlation between the total number of GCs in a galaxy and its halo virial mass.} However, it is very likely that the number of retrograde clusters is larger than the total number detected in this study and that M33 should also have clusters from accretion events that are now in prograde motion. Because the virial mass of M33 halo is $M_{\rm vir} \sim 5 \times 10^{11}$\msun\ \citep{Zacharie2017}, we infer that at least 10\% of the total mass in the halo of M33 has an accreted origin. M33 has a larger halo mass than the Large Magellanic Cloud (LMC), and the MCs have been accreted onto the MW with 70$^{+30}_{-40}$ satellite galaxies \citep{Drlica-Wagner2015,Jethwa2016}. Consequently, M33 has probably hosted tens of satellites over the course of its lifetime. In this study, we have identified the potential residuals left behind in M33's halo by some of its former satellite galaxies.

\section*{Acknowledgements}
BA and GFL thank the organizers of the ``Wide-Field Spectroscopy vs. Galaxy Formation Theory'' meeting held at BioSphere 2, Arizona in March 2023. BA also thanks Julianne Dalcanton for useful comments and suggestion during a visit to the Flatiron Institute’s Center for Computational Astrophysics (CCA) in New York City in November 2023. BA warmly thanks Helios Anguiano D\'enat for his inspiration and the joy he brought during the completion of this work. The authors thank the anonymous referee for suggestions and comments that improved the manuscript. This research has used the SIMBAD database, operated at CDS, Strasbourg, France. This work presents results from the European Space Agency (ESA) space mission Gaia. Gaia data are being processed by the Gaia Data Processing and Analysis Consortium (DPAC). Funding for the DPAC is provided by national institutions, in particular the institutions participating in the Gaia MultiLateral Agreement (MLA). The Gaia mission website is https://www.cosmos.esa.int/gaia. The Gaia archive website is https://archives.esac.esa.int/gaia. This research has made use of NASA's Astrophysics Data System Bibliographic Services. This project has been supported by the Spanish Ministry of Science, Innovation and Universities and the State Research Agency (MICIU/AEI) with the grant RYC2022-037011-I and by the European Social Fund Plus (FSE+).

\section*{Data Availability}

The authors declare that all data used in this work are available from the literature \citep{chandar+2002,sharina+2010,beasley+2015}.



\bibliographystyle{mnras}
\bibliography{literature} 




\appendix




\bsp	
\label{lastpage}
\end{document}